\documentclass[mathleft
]{an}
\usepackage{graphicx}
\usepackage{times}
\def\sn{\hbox{S/N}}

\def\kms{\hbox{km\,s$^{-1}$}}

\def\em{\it}  
  
\def\degr{\hbox{$^\circ$}}

\def\kis{\hbox{$\chi^2$}}

\begin{document}

\Pagespan{789}{}
\Yearpublication{2011}%
\Yearsubmission{2011}%
\Month{11}%
\Volume{999}%
\Issue{88}%

\title{Weak magnetic fields of intermediate-mass stars\thanks{Based on observations obtained at the Bernard Lyot Telescope (TBL, Pic du Midi, France) of the Midi-Pyr\'en\'ees Observatory, which is operated by the Institut National des Sciences de l'Univers of the Centre National de la Recherche Scientifique of France, and at the Canada-France-Hawaii Telescope (CFHT) which is operated by the National Research Council of Canada, the Institut National des Sciences de lÕUnivers of the Centre National de la Recherche Scientifique of France, and the University of Hawaii.}}

\author{P. Petit\inst{1,2}\fnmsep\thanks{Corresponding author:
  \email{ppetit@irap.omp.eu}\newline}
\and  F. Ligni\`eres\inst{1,2}
\and  G.A. Wade\inst{3}
\and  M. Auri\`ere\inst{1,2}
\and  D. Alina\inst{1,2}
\and  T. B\"ohm\inst{1,2}
\and  A. Oza\inst{1,2}
}
\titlerunning{Weak magnetic fields of intermediate-mass stars}
\authorrunning{P. Petit et al.}
\institute{
Universit\'e de Toulouse, UPS-OMP, Institut de Recherche en Astrophysique et Plan\'etologie, Toulouse, France
\and 
CNRS, Institut de Recherche en Astrophysique et Plan\'etologie, 14 Avenue Edouard Belin, F-31400, Toulouse, France
\and
Department of Physics, Royal Military College of Canada, PO Box 17000, Station Forces, Kingston, Ontario, Canada
}

\received{}
\accepted{}
\publonline{}


\abstract{We present the result of a highly sensitive spectropolarimetric study dedicated to intermediate mass stars. We report the detection of sub-gauss surface magnetic fields on the normal, rapidly-rotating A-type star Vega and on the moderately-rotating Am star Sirius A. These magnetic detections constitute the first evidence that tepid stars that do not belong to the class of Ap/Bp stars can also host magnetized photospheres, suggesting that a significant fraction of stars in this mass regime are magnetic. We present here the observational clues gathered so far to progress towards understanding the physical processes at the origin of this newly identified Vega-like magnetism.}

\maketitle

\section{Introduction}

\subsection{Magnetic dichotomy in tepid stars}

\begin{figure*}
\centering
\mbox{
\includegraphics[width=17cm]{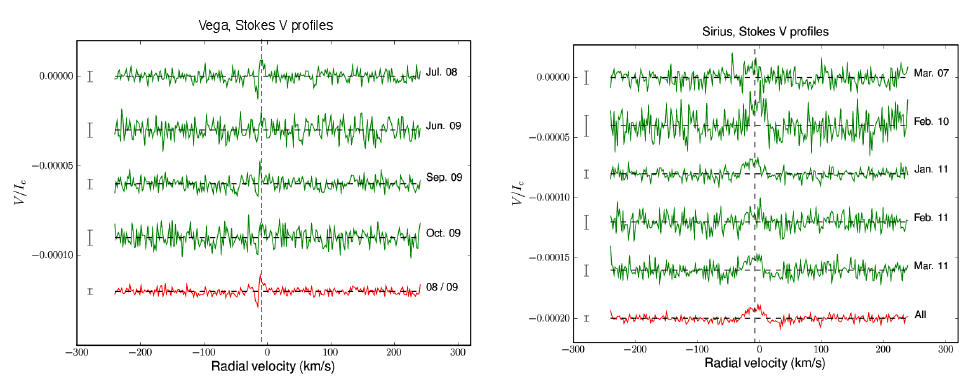}
}
\caption{Averaged Stokes V LSD profiles of Vega (left) and Sirius A (right) for the various observing runs (green lines). The red line is obtained by averaging all Stokes V profiles. Note that successive profiles are shifted vertically for better clarity. The error bar corresponding to each profile is plotted at left. The dashed, vertical lines depict the mean radial velocity of the star. After Petit et al. (2010, 2011).}
\label{fig:stokesv}
\end{figure*}

Among main-sequence stars of late-B and early-A spectral types (the so-called ``tepid'' stars), about 5 to 10 \% display strong magnetic fields at the photospheric level. Since the first stellar magnetic detection of Babcock (1947) in the A1 star 78~Vir, all strongly magnetic stars identified amid tepid stars also belong to the class of chemically-peculiar, Ap/Bp stars. Reciprocally, magnetic fields are detected in all Ap stars observed with sufficient polarimetric accuracy, outlining a minimum longitudinal field strength of about 100~G (Auri\`ere et al. 2007). The surface magnetic topology of Ap stars is very stable with time and is organized in a simple geometry, generally dominated by a dipolar structure (e.g. L\"uftinger et al. 2010), coexisting with smaller magnetic spots that show up in tomographic models based on the four Stokes parameters (Kochukhov et al. 2004, Kochukhov \& Wade 2010). Systematic searches for magnetic fields in intermediate-mass stars using high-resolution spectropolarimetry have failed to discover magnetic objects outside the class of Ap/Bp stars, in spite of longitudinal field accuracy of a few gauss (Shorlin et al. 2002, Wade et al. 2006, Auri\`ere et al. 2010, Makaganiuk et al. 2011), progressively confirming the simple picture of a magnetic dichotomy in tepid and massive stars.

As a tentative interpretation of this magnetic segregation, Auri\`ere et al. (2007) have proposed that a kink-type instability may be responsible for the destruction of large-scale magnetic fields in most A stars. This process may settle whenever a significant differential rotation is operating in the star, which can happen if Maxwell stresses are not high enough to inhibit large-scale plasma motions. Auri\`ere et al. (2007) derived an order of magnitude estimate of the critical magnetic field strength, below which Lorentz forces should no longer be able to enforce a solid-body rotation. The value obtained is in agreement with the minimal field observed in Ap stars and also predicts that the critical magnetic field is proportional to the star rotation rate. When the large scale field configuration is unstable, the observed field is expected to be small due to the cancellation effect of opposed polarities on the (disk-integrated) longitudinal field estimate. The numerous upper limits obtained on magnetic fields in HgMn, Am and normal tepid stars suggest that a sub-gauss detection threshold is a minimal requirement to probe this elusive magnetic population.

\subsection{Vega-like magnetism}

The detection of weak Zeeman signatures formed in the photosphere of the normal, rapidly-rotating A-type star Vega was the first detection of a sub-gauss magnetic field on a main-sequence star of intermediate mass (Ligni\`eres et al. 2009). This discovery was later followed by the detection of polarized signatures on the Am star Sirius A (Petit et al. 2011), suggesting that Vega-like magnetism may be a widespread property of stars in this mass domain. Confronted with what might be a new type of stellar magnetism, we aim at gathering more observational clues about the origin of these weak surface fields and providing theoreticians with observational constraints that can guide them towards identifying and modelling the physical ingredients at the origin of Vega-like magnetism. 

We summarize here the main results obtained in this project. We first detail the technical challenges related to the detection of sub-gauss magnetic fields in tepid and massive stars. Afterwards, we present the main outcome of ultra-deep spectropolarimetric observations of Vega and Sirius A. We finally discuss various options proposed to explain the physical origin of Vega-like magnetic fields.

\section{Utltra-deep spectropolarimetric observations of early A-type stars}

\subsection{Observing material}

The basic material used in this highly-sensitive investigation of weak magnetic fields consists of intensity (Stokes I) and circularly polarized (Stokes V) spectra. We use these data in a search for weak Zeeman signatures. We employ here the NARVAL and ESPaDOnS spectropolarimeters (Auri\`ere 2003) and benefit from their high spectral resolution (close to 65,000), enabling us to comfortably resolve line profiles. Depending on the target, the adopted integration time is chosen to reach peak signal-to-noise ratios (\sn\ hereafter) of about 2,000, which is close to (though safely away from) the saturation threshold of the CCD detector. 

To increase further our capability to detect very weak surface magnetic fields, we use the fact that their polarized signatures are present in all photospheric atomic lines possessing an effective Land\'e factor greater than zero. This multi-line approach is performed using the Least-Squares Deconvolution (LSD) cross-correlation procedure (Donati et al. 1997, Kochukhov et al. 2010). For a main-sequence star of early-A spectral type, about 1,000 lines can be combined together to calculate a mean line profile (LSD profile hereafter) with a typical \sn\ of 20,000.

With the final aim to reach uncertainties as low as 0.1~G in longitudinal field measurements, the last step consists in repeating the whole procedure for a large number of spectra and averaging all LSD profiles together. Using 500 successive observations of the same star, the noise is further decreased by a factor of 25, with a final \sn\ of the grand average reaching a value of up to 500,000. Because such a large number of spectra has to be collected over several nights, the resulting data set contains observations taken at different rotational phases, so that our ultimate LSD profile is providing us with a phase-averaged Zeeman signature. If the signatures of small-scale, non-axisymmetric magnetic features are likely to be averaged out by a significant rotational smearing, this strategy has proved to be a great help in the difficult task of unveiling sub-gauss fields in tepid stars.

\subsection{The normal A-type star Vega}

\begin{figure}
\centering
\mbox{
\includegraphics[width=8cm]{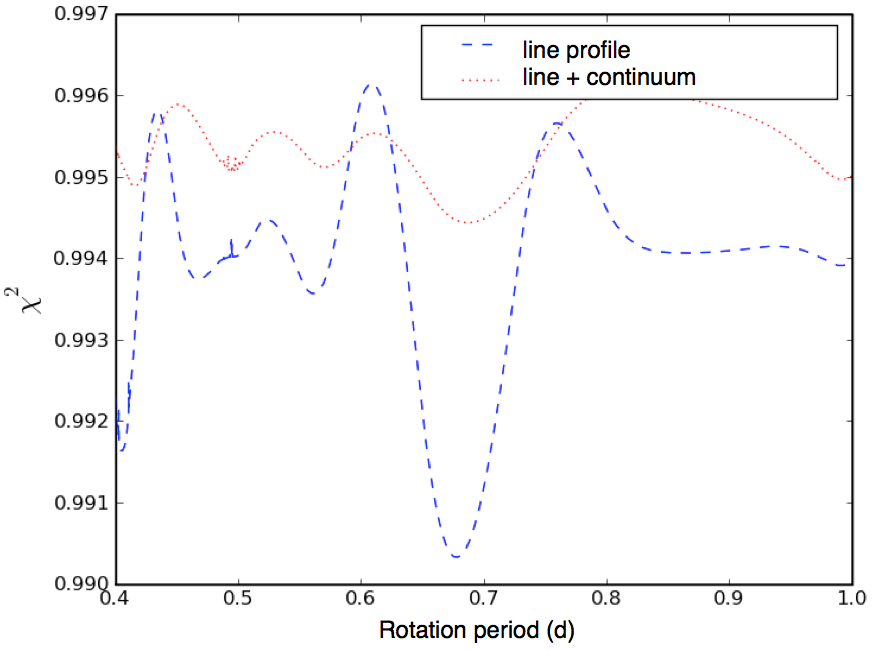}
}
\caption{\kis\ fluctuations, as a function of the rotation period, resulting from a least-squares sine fit to the time-series of 500 Vega Stokes V profiles taken in 2010. The dashed blue curve is obtained over the radial velocity span of the line in the Stokes V LSD profiles. The dotted red line also includes the continuum around the line.}
\label{fig:period}
\end{figure}

This approach was first applied to Vega, a main-sequence A0 star with no strong chemical peculiarities. Evidence for gravity darkening affecting the shape of photospheric lines indicates that Vega is an oblate, rapid rotator seen almost pole-on, with an inclination angle of about 7\degr\ (Takeda et al. 2008), so that its projected rotational velocity of only 20 \kms\ is hiding a much faster equatorial velocity of 170 \kms.

A total of about 1,500 spectra of Vega were recorded with NARVAL and ESPaDOnS, from 2008 to 2011 (Ligni\`eres et al. 2009, Petit et al. 2010, Alina et al. 2011), in an observing effort initially designed to track low-amplitude pulsations on this target (B\"ohm et al. 2011). Averaged Stokes V profiles from the observing campaigns of 2008 and 2009 are plotted in the left panel of Fig. \ref{fig:stokesv}, illustrating the repeated detection of Zeeman signatures at the radial velocity of the star. No significant evolution of the signature is observed over the years, with a phase-averaged, line-of-sight magnetic field component of $-0.6 \pm 0.2$~G (obtained when grouping together all data from 2008 and 2009).

The various time-series gathered for Vega were used to search for a rotational modulation of the polarized signatures. A first method consists in performing a simple least squares sine fit period search to each pixel within the LSD Stokes V profile, for a period range between 0.4 and 1 day, over the time-series of observations (Alina et al. 2011). Using our denser data set (obtained in 2010), a $\chi^2$ minimum is obtained for a period of $0.678^{+0.036}_{-0.029}$~d (Fig. \ref{fig:period}), with consistent estimates derived from our data from 2008 and 2009. A second, more indirect method was carried out using the Zeeman Doppler Imaging technique (ZDI, Donati \& Brown 1997, Donati et al. 2006), in a search for the rotation period that minimizes the $\chi^2$ of the ZDI model fit to the data. Using this tomographic approach, several $\chi^2$ minima of similar depths are generally obtained between 0.4~d and 1~d. The data sets of 2008 and 2009 provide us with a rotation period of $0.732 \pm 0.008$~d (Petit et al. 2010), while a $\chi^2$ minimum close to 0.68~d is derived from the observations of 2010 (Alina et al. 2011).

\begin{figure*}
\centering
\mbox{
\includegraphics[width=4.5cm]{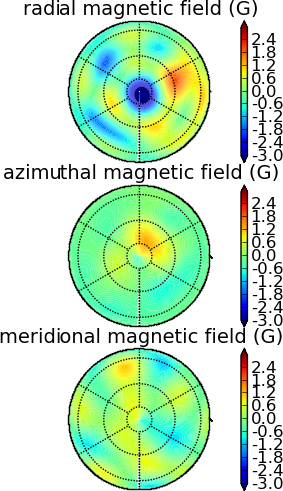}
\hspace{0.5cm}
\includegraphics[width=4.5cm]{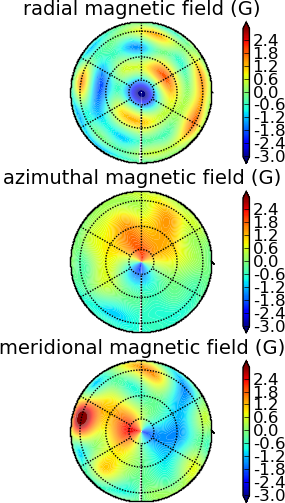}
\hspace{0.5cm}
\includegraphics[width=5.5cm]{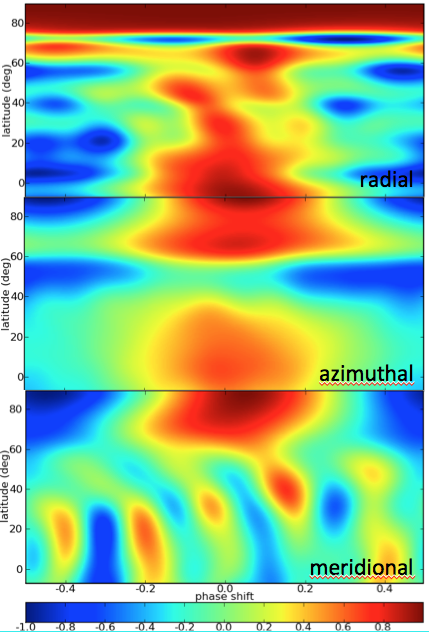}
}
\caption{Left and middle panels : vectorial magnetic maps of Vega for 2008 and 2009, respectivel. A polar  view is adopted. For each map, the 3 vertical charts illustrate the field projection onto one axis of the spherical coordinate frame with the radial, azimuthal, and meridional field components. The magnetic field strength is expressed in gauss. The phase origin is set at the bottom of each chart and rotational phases are increasing in the clockwise direction. Right panel~: cross-correlation maps obtained from the comparison of the magnetic geometries reconstructed in 2008 and 2009. After Petit et al. (2010, 2011).}
\label{fig:maps}
\end{figure*}

Using the ZDI method, two magnetic maps of Vega were reconstructed for 2008 and 2009 (Fig. \ref{fig:maps}), assuming a rotation period of 0.732~d. Because of their high relative noise, the polarized line profiles cannot offer more than weak constraints on the magnetic topology, a limitation that is probably responsible for most of the differences observed between the two maps. In spite of this restriction, several surface characteristics are consistently derived in 2008 and 2009. The most recognizable magnetic feature is the polar spot of radially-oriented field showing up at both epochs. This spatially-limited magnetic region, along with other magnetic spots at lower latitudes, is more complex than the magnetic geometries usually observed on Ap stars, with more than half of the surface magnetic energy of Vega reconstructed in spherical harmonics modes with $\ell > 3$ (Petit et al. 2010). 

Using a cross-correlation procedure to compare the two maps (Fig. \ref{fig:maps}), we find that the highest correlation is achieved for a phase shift close to zero (except at intermediate latitudes in the azimuthal field and at low latitudes in the meridional field), confirming that the maps computed from two independent data sets carry similar information. The absence of any significant phase shift between both epochs suggests that the surface differential rotation, if any, is probably very weak on Vega, so that the photospheric magnetic structure was not significantly distorted by a latitudinal shear within 1 year. 

\subsection{The bright Am star Sirius A}

A similar observing strategy was applied to the weakly metal-rich star Sirius A (Petit et al. 2011). With a surface effective temperature of 9,900~K, Sirius A is among the hottest Am stars. Weak spectral signatures of micro-turbulence indicate the presence of a thin, subsurface convective shell (Landstreet et al. 2009). An interesting difference with Vega is the slower rotation of Sirius A. While its rotation period is unknown, its equatorial velocity should be less than 120~\kms, as slow rotation is a universal property of Am stars (Abt 2009).
  
A total of 442 spectra was collected between 2007 and 2011, using both NARVAL and ESPaDOnS spectropolarimeters. The sets of averaged LSD profiles are plotted in the right panel of Fig.~\ref{fig:stokesv}. Circularly-polarized signatures are consistently detected in line-profiles and are likely generated, through the Zeeman effect, by a weak photospheric magnetic field. Whereas Zeeman signatures extracted from Vega observations show up at the very core of the line (attributed to the polar concentration of the magnetic fied), the signatures detected for Sirius A are broader, which may indicate a global distribution of magnetic regions, and a possibly simpler magnetic topology. 

Combined with the spectropolarimetric study of Vega, the detection of a magnetic field on Sirius A suggests that sub-gauss surface fields are not the prerogative of rapidly rotating tepid stars and may therefore concern a significant fraction of objects in this mass domain.

\section{Origin of Vega-like magnetism}

\subsection{Fossil field}

Most of the volume of tepid stars is constituted of radiative layers, in which large-scale magnetic fields may survive during a significant fraction of a star's evolution (Moss 2001). The origin of such fossil fields must be sought during stellar formation. It may be the heritage of the field threading the molecular cloud that gave birth to the star, or the frozen remnant of dynamo action that occurred during the pre-main-sequence evolutionary phase. The complexity of the magnetic topology of Vega is at odds with the fossil field hypothesis, since only low-order field geometries are expected to survive over long timescales. Vega is still a young star (a few hundreds of Myr), so that some level of complexity may have survived in its surface magnetic field. We note, however, that some strongly magnetic, much younger tepid stars belonging to the class of Herbig stars (no older than a few Myr) already possess a simple surface magnetic structure (Alecian et al. 2008), confirming that the field topology we reconstruct is quite different from anything observed so far on intermediate-mass stars. 

\subsection{Sub-surface dynamo}

The detection of microturbulent broadening and asymmetries in the line-profiles of tepid stars, up to an effective temperature of  10,000~K, reveals local velocity fields in their shallow convective envelopes (Landstreet et al. 2009). If, in principle, a dynamo could take place in this sub-surface shell, both stars observed so far belong to the upper temperature boundary above which the solar-type surface convection seems to be suppressed. For Sirius, a simple estimate of the available convective flux suggests that the convective energy is too weak to account for the detected magnetic field (Petit et al. 2011). The situation is more complex with Vega, for which the equator could be cooler than the pole by as much as 1,000~K (Takeda et al. 2008), so that convection is likely to be mixing the surface material in the coolest part of the photosphere, taking the shape of a low-latitude convective belt. The main magnetic spot observed on Vega is, however, located at the pole, where a surface temperature close to 10,000~K should prevent any convective flows to settle. In both cases, the option of a sub-surface dynamo is therefore unlikey to provide any acceptable justification for the magnetic detections.

\subsection{Core dynamo}

The core of the star is probably a better place to operate a dynamo, and numerical simulations suggests that a vigorous field generation can be obtain in the convective core of tepid stars (Brun et al. 2005). The photospheric observation of the associated magnetic field would imply that the flux tubes generated in the core have crossed the whole radiative zone, up to the surface. In the case of weakly magnetized flux tubes, the buoyant rise time becomes, however, much longer than the ages of Sirius or Vega (Moss 2001, MacGregor \& Cassinelli 2003, MacDonald \& Mullan 2004, Mullan \& MacDonald 2005). It is therefore unlikely to see the product of such a dynamo reach the photosphere so soon in the life of the star.

\subsection{Dynamo in the iron convection zone}

Tepid stars possess yet another zone in which convective mixing is taking place, because of an opacity peak of iron (Richard et al. 2001), and the Fe convection zone may be the location of a dynamo. In massive stars where this convective region may get close to the surface, Cantiello \& Braithwaite (2011) argue that it could produce detectable surface magnetic fields. In intermediate-mass stars, the Fe convection zone is buried much deeper. Whether flux tubes can rise up to the surface and produce detectable magnetic field remains to be investigated.

\subsection{Dynamo action in radiative layers}

Another option would be the onset of a dynamo in the radiative zone. Several theoretical studies propose that a field can be amplified in these stable layers, although they do not agree on the basic ingredients of the dynamo loop (Spruit 2002, Zahn et al. 2007). Since this type of dynamo could be active close to the surface, the field emergence time is not a problem. It is also expected that the field is better amplified close to the rotation axis, which agrees well with the polar field concentration observed on Vega (see also R\"udiger et al. 2011). For these reasons, a dynamo in the radiative layers offers an attractive interpretation of the observed Zeeman signatures. 

\section{Perspective}

Ultra-deep spectropolarimetric observations of a larger stellar sample will be critical to determine whether the dependence of Vega-like magnetism on various stellar parameters (mass, rotation, age or metallicity) meets the expectations of one or several of the theoretical frameworks evoked above. This extended survey is now under way. We are also engaged in a long-term monitoring of Vega, in order to investigate the stability of its magnetic geometry, and to determine whether its magnetic variability, if any, can be related to specific large-scale surface flows, like latitudinal differential rotation or meridional motions.

\acknowledgements

We are grateful to the service observers and technical staffs of TBL and CFHT, for their help in collecting so many photons of stars much brighter than the usual targets of NARVAL and ESPaDOnS.


\end{document}